# Single-Shot 3D Diffractive Imaging of Core-Shell Nanoparticles with Elemental Specificity

Alan Pryor, Jr[1], Arjun Rana[1], Rui Xu[1], Jose A. Rodriguez[2], Yongsoo Yang[1], Marcus Gallagher-Jones[1], Huaidong Jiang[3], Jaehyun Park[4,5], Sunam Kim[4,5], Sangsoo Kim[4,5], Daewong Nam[4,6], Yu Yue[7], Jiadong Fan[3], Zhibin Sun[3], Bosheng Zhang[8], Dennis F. Gardner[8], Carlos Sato Baraldi Dias[1], Yasumasa Joti[9], Takaki Hatsui[4], Takashi Kameshima[9], Yuichi Inubushi[9] Kensuke Tono[9], Jim Yang Lee[7], Makina Yabashi[4], Changyong Song[4,6], Tetsuya Ishikawa[4], Henry C. Kapteyn[8], Margaret M. Murnane[8], Jianwei Miao[1*]

[1]Department of Physics & Astronomy and California NanoSystems Institute, University of California, Los Angeles, CA, 90095, USA

[2]Department of Biological Chemistry, UCLA-DOE Institute for Genomics and Proteomics, University of California, Los Angeles, CA, 90095, USA

[3]School of Physical Science and Technology, ShanghaiTech University, Shanghai 201210, China

[4]RIKEN SPring-8 Center, Kouto 1-1-1, Sayo, Hyogo 679-5148, Japan

[5]Pohang Accelerator Laboratory, Pohang 790-784, South Korea

[6]Department of Physics, Pohang University of Science and Technology, Pohang 790-784, South Korea

[7]Department of Chemical and Biomolecular Engineering, National University of Singapore, 10 Kent Ridge Crescent, Singapore 119260

[8]JILA, University of Colorado at Boulder, Boulder, CO 80309, USA

[9]Japan Synchrotron Radiation Research Institute, Kouto 1-1-1, Sayo, Hyogo 679-5198, Japan

**We report 3D coherent diffractive imaging of Au/Pd core-shell nanoparticles with 6 nm resolution on 5-6 femtosecond timescales. We measured single-shot diffraction patterns**

of core-shell nanoparticles using very intense and short x-ray free electron laser pulses. By taking advantage of the curvature of the Ewald sphere and the symmetry of the nanoparticle, we reconstructed the 3D electron density of 34 core-shell structures from single-shot diffraction patterns. We determined the size of the Au core and the thickness of the Pd shell to be 65.0±1.0 nm and 4.0±0.5 nm, respectively, and identified the 3D elemental distribution inside the nanoparticles with an accuracy better than 2%. We anticipate this method can be used for quantitative 3D imaging of symmetrical nanostructures and virus particles.

Core-shell nanoparticles exhibit unique electronic, chemical, catalytic and optical properties that have found applications across several disciplines [1-4]. Conventional methods to characterize these nanoparticles rely on electron microscopy, scanning probe microscopy, x-ray diffraction, scattering and spectroscopic techniques. Although atomic electron tomography has recently been used to determine the 3D structure of nanoparticles at the single atom level, the sample has to be thin enough to mitigate the dynamical scattering effect [5,6]. Scanning probe microscopy is limited to studies of surface structures and x-ray diffraction and scattering methods only provide average structural information [1,3,4]. In contrast, coherent diffractive imaging (CDI) can be used to determine the 3D internal electron density of thick samples at high resolution [7-9]. Following the first experimental demonstration in 1999 [10], CDI and its variants have been successfully applied to image a broad range of samples in physics, chemistry, materials science, nanoscience and biology [7-9,11-22].

With the advent of x-ray free electron lasers (XFELs) that produce extremely intense and short x-ray pulses [7,23,24], CDI has opened the door for high-resolution imaging of both physical and biological specimens based on the diffraction-before-destruction scheme [25,26]. However, because an intense XFEL pulse will destroy a specimen after one exposure, it would be desirable to find a way to obtain 3D structure information from a single x-ray pulse. One method to achieve single-shot 3D structure determination is the use of the curvature of Ewald sphere and additional constraints such as symmetry and sparsity [27-30]. Here, we report for the first time quantitative single-shot 3D imaging of symmetrical nanoparticles with elemental specificity. We reconstructed the 3D electron density of individual Au/Pd core-shell nanoparticles from single-shot diffraction patterns with 6 nm resolution, and also determined the size of the Au core and the thickness of the Pd shell to be 65.0±1.0 nm and 4.0±0.5 nm, respectively. We quantified the 3D elemental distribution inside the nanoparticle with an accuracy better than 2%. Finally, by implementing a semi-automated data analysis and 3D reconstruction pipeline, we established a general method for high-throughput, quantitative characterization of a wide range of symmetrical nanostructures.

Au/Pd core-shell nanoparticles were synthesized by a seed mediated growth method [31]. Au nanoparticles with truncated cubic shapes were prepared first as the cores. After the epitaxial growth of a Pd shell on the cubic Au core, the composite nanoparticles adopted a perfect cubic shape. SEM and TEM images shows a monodisperse shape and size distribution of the nanoparticles (Fig. 1 insets). The formation of Au/Pd core-shell structure was also implicated by the alternating bright and dark fringes in the TEM image caused by the superposition of two misfit crystalline lattices in a core-shell construction. The XFEL experiment was conducted using the SPring-8 Angstrom Compact Free Electron Laser [24]. Figure 1 shows the schematic layout of the single-shot 3D diffractive imaging experiment. X-ray pulses with an energy of 6 keV and a repetition rate of 10 Hz were focused to a 1.5 µm spot by a pair of Kirkpatrick-Baez (K-B) mirror. Each pulse contained ~$10^{11}$ photons with a pulse duration of 5-6 fs (Fig. S1 in the supplemental materials [32]). Nanoparticles were deposited onto a 100-nm-thick $Si_3N_4$ membrane grid and inserted into a multi-application x-ray imaging chamber where the sample was scanned relative to the x-ray pulses [33]. Single-shot x-ray diffraction patterns were measured by an octal multi-port charge-coupled device [34]. The nanoparticles were destroyed after the impinging of x-ray pulses, leaving small holes on the $Si_3N_4$ membrane (Fig. 1 inset). A total of 39,151 diffraction patterns were acquired consisting of no, partial, single, and multiple hits. Single-hits were separated from these patterns based on a threshold of the average diffraction intensity. A good hit rate for single-shot diffraction pattern is around a few percent. From the hit candidates, we selected a subset of 34 diffraction patterns for further analysis.

The 34 diffraction patterns were processed and reconstructed by using a semi-automated 3D data analysis pipeline, shown in Fig. 2. From each diffraction pattern, the background was subtracted based on the most recently available background exposure. An additional flat background subtraction was required, the value of which was determined by first smoothing and thresholding each pattern to determine the background region. The final subtracted value was determined by the average nonnegative pixel intensity in the background region multiplied by a single scaling factor, whose value was optimized based upon the quality of the resulting reconstructions. The center of each diffraction pattern was determined based on the centro-symmetry of the diffraction intensity at the low spatial frequency. Since the diffraction patterns have larger oversampling ratios [35], each pattern was binned by 9×9 pixels to enhance the signal-to-noise ratio [36]. Figure S2 in the supplemental materials [32] shows the 34 processed single-shot diffraction patterns, in which the diffraction signal is limited by the size of the detector. The orientation of each single-shot diffraction pattern can in principle be determined by the self-common arc method [30]. But the 34 diffraction patterns in this experiment were oriented close to the four-fold symmetry axis as the majority of nanocubes sit flat on the surface of $Si_3N_4$ membranes (Fig. 1 insets). This allowed us to develop a simpler approach to refine the orientation of each diffraction pattern. We first estimated the size of a nanocube based on the speckle size and experimental parameters. We then slightly changed the orientation of the nanocube and calculated the corresponding diffraction pattern. By minimizing the difference between the calculated and measured diffraction patterns, we determined the orientation of each diffraction pattern with an angular precision of ~0.5°.

Each diffraction pattern was then projected onto the surface of the Ewald sphere [27]. By taking into account of the curvature of the Ewald sphere and the 48 octahedral symmetry operations, a 3D Cartesian grid of the Fourier magnitudes was assembled by the following interpolation approach,

$$|F_{obs}(\vec{k})| = \frac{\Sigma_i \frac{|F(\vec{k}_i)|W(\vec{k}_i)}{\Delta\Omega_i|\vec{k}_i-\vec{k}|}}{\Sigma_i \frac{W(\vec{k}_i)}{|\vec{k}_i-\vec{k}|}} \quad ; \quad W(\vec{k}_i) = \begin{cases} 1, & \frac{|\vec{k}_i-\vec{k}|}{\Delta p} < d_c \\ 0, & \frac{|\vec{k}_i-\vec{k}|}{\Delta p} \geq d_c \end{cases} \quad (1)$$

where $|F_{obs}(\vec{k})|$ is the interpolated Fourier magnitudes on the 3D Cartesian grid, $|F(\vec{k}_i)|$ is the measured Fourier magnitudes of the $i^{th}$ pixel projected onto the surface of the Ewald sphere, $W(\vec{k}_i)$ represents a spherical interpolation kernel of radius $d_c$ (where $d_c$ = 0.7 voxels in this case), $\Delta\Omega_i$ is the solid angle subtended by the $i^{th}$ pixel of the detector and $\Delta p$ is the pixel size in reciprocal space. When the diffraction pattern has a large oversampling ratio and the Fourier magnitudes change smoothly [35], this interpolation approach is computationally efficient and accurate.

Using Eq. (1), we produced a 3D Cartesian grid of the Fourier magnitudes for each single-shot diffraction pattern. A fraction of the grid points were filled in by the measured data and the remaining points were set as undefined. The phase retrieval was carried out by the oversampling smoothness (OSS) algorithm [37]. A total of 1,000 independent, randomly seeded 3D reconstructions were performed for each 3D grid of the Fourier magnitudes. Each reconstruction consisted of 1,000 iterations of OSS with ten progressive filters, positivity constraint, and a loose support. The algorithm iterated between real and reciprocal space. The positivity and support constraints were applied in real space and the measured grid points were enforced in reciprocal space, while the undefined points were iteratively determined by the algorithm. An R-factor, defined as the sum of difference between measured and calculated Fourier magnitudes normalized by the sum of measured Fourier magnitudes, was used to monitor the convergence of the iterative algorithm. After 1,000 iteration, the majority of 1,000 independent reconstructions had converged and the top 10% with the smallest R-factors were averaged to obtain a final 3D reconstruction. Because the quantity of data obtained during an XFEL experiment is so high, we have implemented a semi-automated pipeline for diffraction pattern selection, data analysis and 3D reconstruction (Fig. 2), allowing the visualization of the final 3D reconstructions during the experiment. By using this semi-automated pipeline, we obtained the final reconstructions of the 34 single-shot diffraction patterns (Fig. S3 in the supplemental materials [32]). Figure 3(a) and movie S1 show the iso-surface renderings of a representative final reconstruction, in which the core and shell structures are clearly visible. To quantify the resolution, we calculated the Fourier shell correlation (FSC) between the final reconstructions of different single-shot diffraction patterns, which has been widely used to estimate the resolution in single-particle cryo-electron microscopy [38]. Based on the criterion of FSC = 0.5, we estimated a 3D resolution of 6 nm was achieved for the reconstructions (Fig. 3(b)). The sudden drop of the FSC curve corresponds to the cut-off of the diffraction intensity by the detector edge, indicating that either the use of a larger detector or shortening the distance between the sample and the detector will improve the resolution.

Next, we developed a model-fitting procedure to quantify the size and electron density of the core-shell nanoparticles. For each of the top 10% independent reconstructions resulting from a single-shot diffraction pattern, a series of model core-shell nanocubes were compared using an error function, defined as

$$Err = \frac{\sum_{\vec{r}} |\rho_{rec}(\vec{r}) - \rho_{mod}(\vec{r})|}{\sum_{\vec{r}} \rho_{mod}(\vec{r})} \qquad (2)$$

where $\rho_{rec}(\vec{r})$ and $\rho_{mod}(\vec{r})$ are the electron intensity of the reconstruction and model, respectively. Each model was created at five times the voxel resolution of the reconstructed nanocubes. The core size, shell thickness, and ratio of core to shell density were varied, and the model was then binned and compared with the reconstruction. The model with the lowest error was recorded for each reconstruction, and the parameters of the matched models were statistically combined to assign a final fit to the set of reconstructions from a single-shot diffraction pattern. The results of this fitting are shown in Fig. 4. The average nanoparticle consists of a 4.0 ± 0.5 nm thick shell of Pd surrounding a uniform 65.0 ± 1.0 nm Au core. The average intensity ratio between the Au core and Pd shell was 1.69, which is within 2% agreement of the tabulated scattering factor ratio of 1.72 [39]. For 3D reconstruction methods that use multiple pulses, interpreting the reconstructed intensity as a direct representation of the 3D electron density is complicated by the inherent instability in the XFEL pulses [23,24], requiring nontrivial normalization before multiple exposures are merged into a single structure. Single-shot 3D reconstruction methods do not suffer from this problem, as the ratio of scattering factors is independent of the overall beam intensity.

The ability to achieve single-shot 3D diffractive imaging of core-shell nanoparticles with elemental specificity has four significant implications. First, although we used core-shell nanocubes as a model system to demonstrate the quantitative characterization ability, this method can be applied to characterize the 3D structure of a wide range of nanoparticles with octahedral, icosahedral, cuboctahedral, decahedral, and trisoctahedral symmetry. Second, a 3D resolution of 6 nm is currently limited by the geometry of the experimental set-up.

Further improvement of the instrumentation and the peak power of the XFEL pulse would allow sub-nanometer 3D resolution to be achieved without sacrificing the field of view of the sample [40]. Third, we have implemented a semi-automated data analysis and 3D reconstruction pipeline for high-throughput quantitative characterization of symmetrical nanostructures using XFELs. This quantitative characterization approach, coupled with advanced synthesis and computational methods such as molecular dynamics and *ab initio* calculations, is expected to expand our understanding of the critical structural and morphological features required to make superior catalysts, adsorbents, electrodes or semiconductors. Finally, we performed numerical simulations on single-shot 3D imaging of virus particles using XFEL pulses (supplemental text and Fig. S4 in the supplemental materials [32]), indicating our method can be used to determine the 3D structure of individual symmetrical viruses without the requirement of averaging identical copies.

In conclusion, we demonstrated quantitative 3D imaging of Au/Pd core-shell nanoparticles with elemental specificity using single XFEL pulses. These core-shell structures are representative of a vast library of nanoparticles with varying chemical, catalytic, optical and electronic properties [1-4]. We developed a semi-automated and quantitative routine for analyzing nanostructures, and applied it to 34 isolated nanoparticles. Using the curvature of the Ewald sphere and symmetry intrinsic to the nanoparticle, we reconstructed highly reproducible 3D structures from single-shot diffraction patterns with a 3D resolution of 6 nm on 5-6 fs timescales. We determined the size of the Au core and the thickness of the Pd shell to be $65.0 \pm 1.0$ nm and $4.0 \pm 0.5$ nm, respectively. The reconstructed electron density of the core and shell structure matches the tabulated scattering factor ratio of Au/Pd within a 2% deviation. This level of accuracy lends itself to quantitative high-resolution 3D characterization of a broad range of symmetric nanostructures in a high-throughput fashion.

This work was supported by the National Science Foundation (NSF) under Grant No. DMR-1548924 and DARPA PULSE program through a grant from AMRDEC. J. M. acknowledges the partial support by NSF (DMR-1437263) and ONR MURI (N00014-14-1-0675). The experiments were performed at the SPring-8 Angstrom Compact Free Electron Laser in Japan (Proposal No. 2013B8014).

**References**

*Email: miao@physics.ucla.edu

1. B. O. Dabbousi et al., J. Phys. Chem. B, **101**, 9463–9475 (1997).

2. S. Alayoglu, A. U. Nilekar, M. Mavrikakis and B. Eichhorn, Nat. Mater. **7**, 333 - 338 (2008).

3. F. Tao et al., Science **322**, 932-934 (2008).

4. R. G. Chaudhuri and S. Paria, Chem. Rev. **112**, 2373–2433 (2012).

5. M. C. Scott *et al*., Nature **483**, 444 (2012).

6. J. Miao, P. Ercius and S. J. L. Billinge, Science, 353, aaf2157 (2016).

7. J. Miao, T. Ishikawa, I. K. Robinson and M. M. Murnane, Science **348**, 530–535 (2015).

8. I. Robinson and R. Harder, Nat. Mater. **8**, 291 - 298 (2009).

9. H. N. Chapman and K. A. Nugent, Nat. Photonics **4**, 833–839 (2010).

10. J. Miao, P. Charalambous, J. Kirz and D. Sayre, Nature **400**, 342-344 (1999).

11. J. Miao et al., Phys. Rev. Lett. **89**, 088303 (2002).

12. D. Shapiro et al., Proc. Natl. Acad. Sci. U.S.A. **102**, 15343–15346 (2005).

13. M. A. Pfeifer, G. J. Williams, I. A. Vartanyants, R. Harder, I. K. Robinson, Nature **442**, 63–66 (2006).

14. H. N. Chapman et al., J. Opt. Soc. Am. A **23**, 1179–1200 (2006).

15. J. M. Rodenburg et al., Phys. Rev. Lett. **98**, 034801 (2007).

16. P. Thibault et al., Science **321**, 379–382 (2008).

17. Nishino, Y. Takahashi, N. Imamoto, T. Ishikawa, K. Maeshima, Phys. Rev. Lett. **102**, 018101 (2009).

18. H. Jiang et al., Proc. Natl. Acad. Sci. U.S.A. **107**, 11234–11239 (2010).

19. M. Dierolf et al., Nature **467**, 436–439 (2010).

20. K. Giewekemeyer et al., Proc. Natl Acad. Sci. USA **107**, 529–534 (2010).

21. A. Rodriguez et al., IUCrJ **2**, 575–583 (2015).

22. A. Ulvestad et al., Science **348**, 1344-1347 (2015).

23. P. Emma et al., Nat. Photonics **4**, 641–647 (2010).

24. T. Ishikawa et al., Nat. Photonics **6**, 540–544 (2012).

25. R. Neutze, R. Wouts, D. van der Spoel, E. Weckert, J. Hajdu, Nature **406**, 752–757 (2000).

26. H. N. Chapman et al., Nat. Phys. **2**, 839–843 (2006).

27. K. S. Raines et al., Nature **463**, 214-217 (2010).

28. C.-C. Chen et al., Phys. Rev. B **84**, 024112 (2011).

29. Maor Mutzafi et al., Nat. Commun. **6**, 7950 (2015).

30. R. Xu et al., Nat. Commun. **5**, 4061 (2014).

31. Y. Yu, Q. Zhang, B. Liu, J. Y. Lee, J. Am. Chem. Soc. **132**, 18258-18265 (2010).

32. See Supplemental Material at http://link.aps.org/XXX for details.

33. C. Song et al., J. Appl. Cryst. 47, 188-197 (2014).

34. T. Kameshima et al., Rev. Sci. Instrum. **85**, 033110 (2014).

35. J. Miao, D. Sayre and H. N. Chapman, J. Opt. Soc. Am. A **15**, 1662-1669 (1998).

36. C. Song et al., Phys. Rev. B **75**, 012102 (2007).

37. J. A. Rodriguez, R. Xu, C.-C. Chen, Y. Zou and J. Miao, J. Appl. Cryst. **46**, 312-318 (2013).

38. J. Frank, *Three-Dimensional Electron Microscopy of Macromolecular Assemblies* (Oxford University Press, 2006).

39. B.L. Henke, E.M. Gullikson, J.C. Davis, Atom. Data Nucl. Data Tabl. **54**, 181-342 (1993).

40. J. Miao, T. Ishikawa, E. H. Anderson and K. O. Hodgson, Phys. Rev. B. **67**, 174104 (2003).

**Figure Captions**

**FIG 1**. Schematic layout of the single-shot 3D diffractive imaging set-up. XFEL pulses with an energy of 6 keV and a pulse duration of 5-6 fs were focused to a 1.5 µm spot by a pair of K-B mirrors. A four-way cross slit was used to eliminate the parasitic scattering from the mirrors. Au/Pd core-shell nanoparticles with a monodisperse shape and size distribution (insets) were supported on a 100-nm-thick $Si_3N_4$ membrane grid and raster scanned relative to the focused beam. Each intense x-ray pulse produced a single-shot diffraction pattern, recorded by an octal multi-port charge-coupled device. A small hole was created on the $Si_3N_4$ membrane after a single exposure (insets).

**FIG 2**. Semi-automated data analysis and 3D reconstruction pipeline. (a) A large number of diffraction patterns were experimentally collected consisting of no, partial, single, and multiple hits by XFEL pulses. High-quality single-hit diffraction patterns were selected from these patterns. The different colors in the pattern are due to the difference of the read-out noise of the detector segments. (b) After background subtraction and center localization, each diffraction pattern was binned by 9×9 pixels to enhance the signal-to-noise ratio and the orientation of the pattern was determined. (c) By taking advantage of the curvature of the Ewald sphere and symmetry intrinsic to the nanoparticle, a single-shot diffraction pattern was used to produce a 3D Cartesian grid of the Fourier magnitudes by a gridding method. (d) The 3D phase retrieval was performed by the OSS algorithm. Among 1,000 independent

reconstructions, the top 10% with the smallest R-factors were averaged to obtain a final 3D reconstruction for each single-shot diffraction pattern.

**FIG 3**. Quantitative analysis of the 3D recosntruction. (a) Iso-surface renderings of a reconstructed core-shell nanoparticle, showing a Au core surrounded by a Pd shell. (b) Average Fourier shell correlation (FSC) between every pair of the 34 reconstructed nanoparticles, indicating a 3D resolution of 6 nm based on the criterion of FSC = 0.5. (c) Central 32-nm-thick slice of a final 3D reconstruction with an overlaid line scan plotted in (d), showing the electron density variation of the Au core and Pd shell.

**FIG 4**. Quantification of the Au/Pd core-shell structure using a model-fitting procedure. (a) and (b) The distribution of the core size and shell thickness obtained from 34 single-shot diffraction patterns. Each data point shows the mean and standard deviation of the top 10% of 1,000 independent reconstructions for a single-shot diffraction pattern. The horizontal red lines indicate the average core size and shell thickness for 34 nanoparticles. (c) and (d) The core/shell distribution of the 34 nanoparticles, indicating the size of the Au core and the thickness of the Pd shell are 65.0±1.0 nm and 4.0±0.5 nm, respectively.

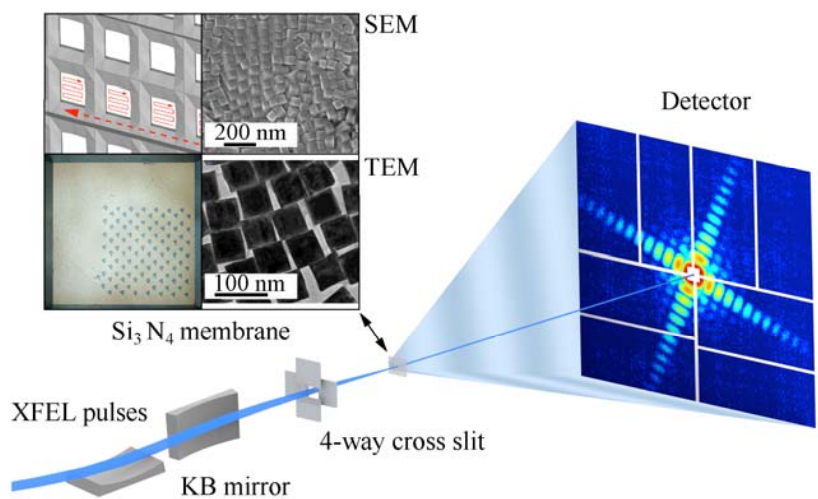

**FIG. 1**

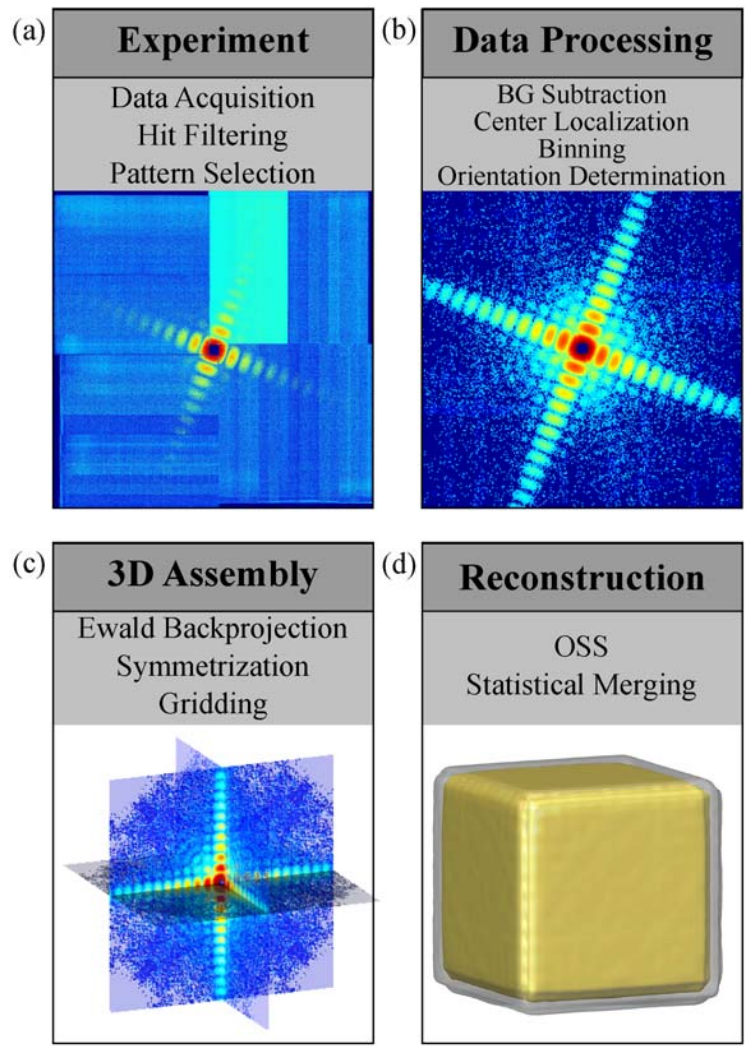

**FIG. 2**

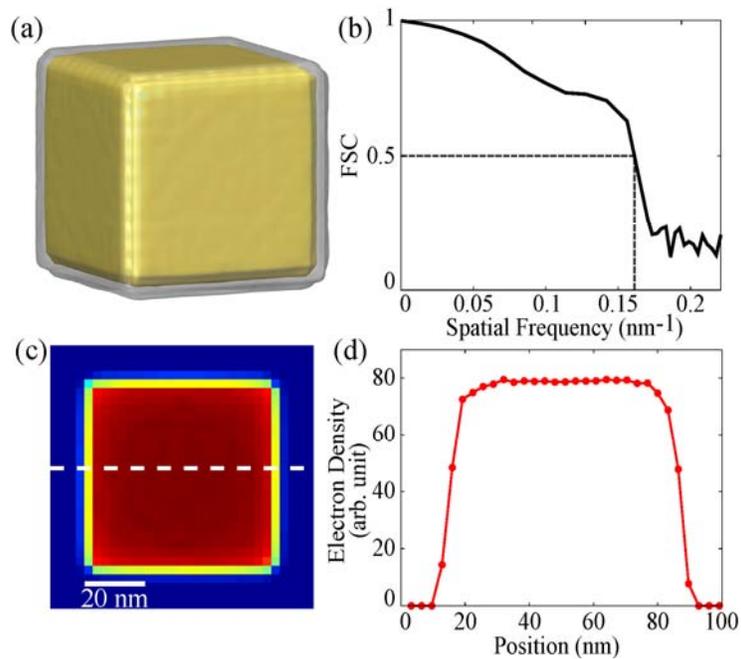

**FIG. 3**

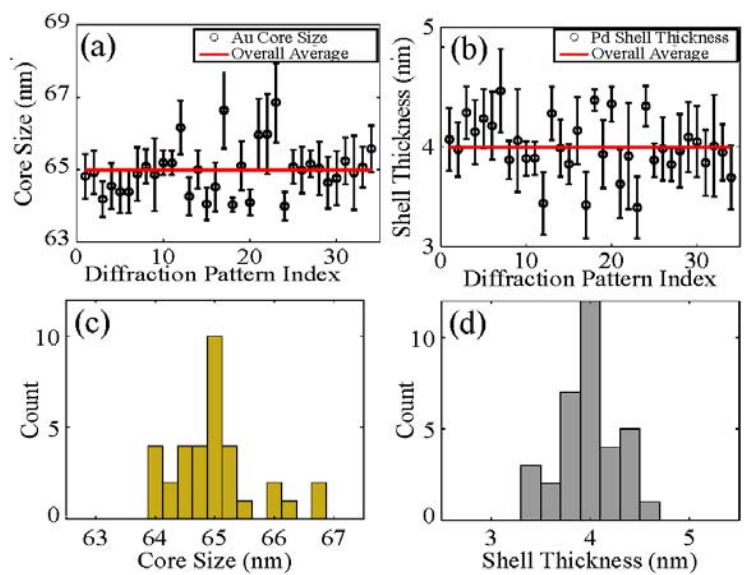

**FIG. 4**